\font\twelverm = cmr10 scaled\magstep1 \font\tenrm = cmr10
       
\font\largerm = cmr10 scaled\magstep2
\font\twelvei = cmmi10 scaled\magstep1
       \font\teni = cmmi10 
\font\twelveit = cmti10 scaled\magstep1 
       
\font\twelvesy = cmsy10 scaled\magstep1
       \font\tensy = cmsy10 
\font\twelvebf = cmbx10 scaled\magstep1 \font\tenbf = cmbx10
       
\font\largebf = cmbx10 scaled\magstep2
\font\twelvesl = cmsl10 scaled\magstep1
\font\twelveit = cmti10 scaled\magstep1
\font\twelvett = cmtt10 scaled\magstep1
\textfont0 = \twelverm \scriptfont0 = \twelverm
       \scriptscriptfont0 = \tenrm
       \def\rm{\fam0 \twelverm}
\textfont1 = \twelvei \scriptfont1 = \twelvei
       \scriptscriptfont1 = \teni
       
\textfont2 = \twelvesy \scriptfont2 = \twelvesy
       \scriptscriptfont2 = \tensy
       
\newfam\itfam \def\it{\fam\itfam \twelveit} \textfont\itfam=\twelveit
\newfam\slfam  \textfont\slfam=\twelvesl
\newfam\bffam \def\bf{\fam\bffam \twelvebf} \textfont\bffam=\twelvebf
       \scriptfont\bffam=\twelvebf \scriptscriptfont\bffam=\tenbf
\newfam\ttfam  \textfont\ttfam=\twelvett
\rm
\hsize=6in
\hoffset=.45in
\vsize=9in
\baselineskip=24pt
\raggedright  \pretolerance = 800  \tolerance = 1100
\raggedbottom
\dimen1=\baselineskip \multiply\dimen1 by 3 \divide\dimen1 by 4
\dimen2=\dimen1 \divide\dimen2 by 2

\nopagenumbers
\headline={\ifnum\pageno=1 \hss\thinspace\hss
     \else\hss\folio\hss \fi}

\count10 = 0
\def\section#1{\vfill\eject \vbox to \dimen1 {\vfill}
    \global\advance\count10 by 1
    \centerline{\uppercase\expandafter{\romannumeral\count10}.\ \bf {#1}}
    \global\count11=96  
    \vskip \dimen1}
\def\subsection#1{\global\advance\count11 by 1
    \vskip \parskip  \vskip \dimen2
    \centerline{{\it {\char\number\count11}\/})\ {\it #1}}
    \vskip \dimen2}
\def\refindent{\advance\leftskip by 24pt \parindent=-24pt}

\parskip=0.2truecm
\def\ltsima{$\; \buildrel < \over \sim \;$}
\def\lsim{\lower.5ex\hbox{\ltsima}}
\def\gtsima{$\; \buildrel > \over \sim \;$}
\def\gsim{\lower.5ex\hbox{\gtsima}}

\def\e{\epsilon}

\def\Ld{{L_{\rm d}}}

\def\Mdot{{\dot M}}
\def\Mdotc{{\dot M_{\rm c}}}
\def\Ms{{M_\odot}}
\def\OmegaK{{\Omega_{\rm K}}}
\def\Qrad{{Q_{\rm rad}^-}}

\def\rS{{r_{\rm g}}}
\def\rg{{r_{\rm g}}}
\def\tff{{\tau_{\rm ff}}}

\def\vp{{v_\varphi}}
\def\vr{{v_{\rm r}}}

\bigskip
\centerline{~~~}
\vskip 0.5in
\centerline{\largebf X-Ray Fluctuations from}
\bigskip
\centerline{\largebf Locally Unstable Advection-Dominated Disks}
\bigskip
\bigskip
\centerline{\largerm
Tadahiro M{\tenrm ANMOTO}$^1$,
Mitsuru T{\tenrm AKEUCHI}$^1$,
Shin M{\tenrm INESHIGE}$^1$,}
\centerline{\largerm
Ryoji M{\tenrm ATSUMOTO}$^2$,
{\rm and}
~Hitoshi N{\tenrm EGORO}$^3$}
\bigskip
\centerline{$^1$
Department of Astronomy, Faculty of Science,
Kyoto University, Sakyo-ku, Kyoto 606-01, Japan}

\centerline{$^2$
Department of Physics, Faculty of Science,
Chiba University, Inage-ku, Chiba 263, Japan}

\centerline{$^3$
Institute of Space and Astronautical Science,
Yoshinodai, Sagamihara 226, Japan}

\bigskip
\bigskip
\bigskip
\centerline{\bf Abstract}

The response of advection-dominated accretion disks to local
disturbances is examined by one-dimensional numerical simulations. It
is generally believed that advection-dominated disks are thermally
stable. We, however, find that any disurbance added onto accretion
flow at large radii does not decay so rapidly that it can move inward
with roughly the free-fall velocity. Although disturbances continue
to be present, the global disk structure will not be modified largely.
This can account for persistent hard X-ray emission with substantial
variations observed in active galactic nuclei and stellar black hole
candidates during the hard state. Moreover, when the disturbance
reaches the innermost parts, an acoustic wave emerges, propagating
outward as a shock wave. The resultant light variation is roughly
(time) symmetric and is quite reminiscent of the observed X-ray shots
of Cygnus X-1.


\vfill\eject

\bigskip
\noindent{\bf 1.~ Introduction}
\medskip

One of the most intriguing features of stellar black-hole candidates (BHCs)
is their rapid X-ray variability or flickering (see e.g., Oda 1979;
Lochner, Swank, \& Szymkowiak 1991; Miyamoto et al.  1991).  It is known
that stellar BHCs, such as Cygnus X-1, have two distinct spectral states;
in the soft (or high) state the emergent spectra are roughly Planckian,
whereas in the hard (or low) state, the spectra are of a power-law type.
Rapid X-ray fluctuations are more pronounced during the hard state
(Miyamoto et al.  1992).  Similar fluctuations exist in active galactic
nuclei, as well (see e.g., Lawrence et al.  1987; McHardy and Czerny 1987).
Their origins need to be understood as a general phenomenon involved in
accretion onto compact objects.

There have been many efforts devoted to the physical understanding of the
origin and mechanism for creating such fluctuations.  (e.g., Terrell
1972,Abramowicz et al.  1991, Mineshige, Takeuchi, \& Nishimori 1994,
Takeuchi, Mineshige, \& Negoro 1995).  Some of the observational features,
such as the power spectra, can be reproduced by these models, but all the
models are still somewhat phenomenological, and the physical cause of the
variabilities remains as one of the great mysteries of X-ray astronomy.

We here stress that the noise characteristics should be understood on the
basis of the disk models.  It is reasonable to assume that the soft-state
disk is the standard-type disk, since it predicts a Planckian spectrum with
a temperature of $\sim$ 1 keV, which is exactly what we observe.  The
standard-type disk is thermally stable, thus producing few fluctuations
during the soft state.  In contrast, the disk structure in the hard state
is poorly understood.  At least, the hard spectrum indicates the disk's
being optically thin.  However, the hard-state disk cannot be of the sort
invoked by Shapiro, Lightman \& Eardley (1976), since this is thermally
unstable (Piran 1978).

Interestingly, there exists a distinct, optically thin solution dominated
by advection (see Narayan \& Yi 1995; Abramowicz et al.  1995).  We
demonstrate here that this type of disk can produce persistent, but
fluctuating hard X-ray emission.  Kato, Abramowicz, \& Chen (1996), in
fact, found by linear analysis that such disks are globally stable, but
could be unstable for short wavelength perturbations.  Since this analysis
is restricted to the linear stage of a thermal instability, its
observational implication are open to question.  We are thus motivated to
perform nonlinear, global numerical simulations of the growths of thermal
perturbations in the optically thin, advection-dominated disks.  The basic
equations and numerical procedures are described in \S 2.  The resultant
time evolution of perturbations (or disturbances) will be presented and
discussed in relation to the observed X-ray variabilities in \S 3.

\bigskip
\noindent{\bf 2.~ Physical Assumptions and Numerical Procedures}
\medskip

Assuming that the disk is geometrically thin, $H < r$, with $H$ and $r$
being the half-thickness and the radius of the disk, we integrate the
disk's structure in the vertical direction.  The basic equations are those
of mass conservation, momentum conservation, angular momentum conservation,
the energy equation (Matsumoto et al.  1984);

$$ {\partial\over\partial t}(r\Sigma)
     + {\partial\over\partial r}(r\Sigma \vr) = 0,
                                                          \eqno(1)$$
$$ {\partial\over\partial t}(r\Sigma \vr)
     + {\partial\over\partial r}(r\Sigma \vr^2)
     = -r{\partial W\over\partial r}
       + r^2\Sigma (\Omega^2 - \OmegaK^2)
       - W{d\ln\OmegaK\over d\ln r},
                                                          \eqno(2)$$
$$ {\partial\over\partial t}(r^2\Sigma \vp)
     + {\partial\over\partial r}(r^2\Sigma \vr \vp)
     = -{\partial\over\partial r}(r^2 \alpha W),
                                                          \eqno(3)$$
and
$$ {\partial\over\partial t}(r\Sigma\e)
     + {\partial\over\partial r}(r\Sigma\e \vr)
     = -{\partial\over\partial r}(rW \vr)
       -{\partial\over\partial r}(r\alpha W \vp) - r\Qrad,
                                                            \eqno(4)$$
where $\Sigma$ is surface density,
      $W$ is integrated pressure,
      $\Omega (=\vp/r)$ and $\OmegaK [=(GM/r)^{1/2}/(r-\rS)$
        with $M$ the mass of the central black hole]
        are the angular frequency of the gas flow
        and Keplerian angular frequency, respectively,
      $\alpha$ is the viscosity parameter,
      $\e$ is the internal energy of accreting gas,
  $\rS \equiv 2GM/c^2 \simeq 3 \times 10^6 (M/10\Ms)$cm
  is the Schwarzschild radius.
  $\Qrad$ is radiative cooling rate per unit surface area
  due to thermal bremsstrahlung.
In the energy equation (4), the second term on the left-hand side
corresponds to advective energy transport, while the last two terms on the
right-hand side represent viscous heating and radiative cooling,
respectively.  Throughout this Letter, we assign $M = 10\Ms$, $\alpha =
0.3$.

\noindent{\it 2.2. ~ Initial and boundary conditions}

We first calculate a steady disk structure by integrating the basic
equations in the radial direction, omitting the terms with time
derivatives.  In the entire region for time-dependent calculations, cooling
is dominated by advection, although we calculate the global steady solution
starting integration at $r=1000\rg$ assuming that $\Omega = \OmegaK$ and
viscous heating is balanced with radiative cooling.  Next, we add a density
perturbation (or disturbance), $\delta\Sigma$, to the steady disk
structure.  Its functional form is ${\delta \Sigma/\Sigma}= 0.2
\exp[-(r-r_0)^2/(\lambda/2)^2]$, where $r_0$ and $\lambda$ are free
parameters, and we set $r_0/\rS = 50.0$ and $\lambda/\rS = 10.0$.  Small
disturbances are added to all physical variables for consistency;
$(\delta\Sigma/\Sigma) \approx kr(\delta W/W)
                       \approx (kr)^{1/2}(\delta v/v)
                       \approx (kr)^{1/2}(\delta\Omega/\Omega)$,
according to the linear analysis (Kato et al.  1996), where $k$ is the
wavenumber of the disturbance, $k = 2\pi/\lambda$.

We add a mass of $\Mdot_0 \triangle t$ into the disk through the outer
boundary at every time step ($\triangle t$).  We assume free boundary
conditions at the inner boundary at $r=2\rg$.  This treatment is
appropriate, since the inner boundary is located in the supersonic region.
Physical quantities at the outer boundary at $r=100\rg$ are set to be equal
to their steady values.

\bigskip
\noindent{\bf 3.~ Time Evolution of Advection-Dominated Disks}
\medskip

\noindent{\it 3.1. ~ Formation of X-ray shots}

We display in Figure 1 the time evolution of an added disturbance in the
three-dimensional $(r, t, \Sigma$) plane.  We assume a mass-input rate of
$\Mdot_0 = 10^{-2}\Mdotc$ with $\Mdotc \equiv 32\pi c\rg/\kappa_{\rm es}$
in the present study.  The lowest curve represents the initial state (a
steady state with a disturbance).  Although the disturbance slowly damps as
it accretes, the damp timescale is comparable to the accretion timescale.
Once a disturbance is given at the outer portions of the disk, that
disturbance will possibly persist as a blob, flowing inward with the same
velocity as that of accreted matter.  The accretion velocity is less than,
but comparable to the free-fall velocity.

According to the linear analysis by Kato et al. (1996), a thermal
instability tends to grow for any perturbation.  The growth timescale is
roughly the same order as the accretion timescale, $\sim (kr)^{-
1/2}(\alpha\Omega)^{-1}$.  The usual viscous diffusion process, on the
other hand, rather suppresses the growth of the perturbations as global
effects.  Therefore, whether the disturbance will grow or damp depends on
the wavelength and nature of the disturbance.  It is important to note that
the disk material will reach the inner edge before the disturbance grows to
significantly alter the global disk structure.  We thus expect persistent
X-ray radiation, although it could be fluctuating at every time.  This is
exactly what is observed in X-ray binaries during the hard (low) state and
in active galactic nuclei.

It is interesting to note that when the disturbance reaches the innermost
parts, it gives rise to an acoustic wave, which propagates outward and
evolves to become a shock wave.  Figure 2 displays the time variations in
the radial profile of Mach number ({\it top}) and that of integrated
pressure ($W$, {\it bottom}).  We have calculated several models with
different $\lambda$-values, finding that the wave reflection occurs only
when $\lambda > \rS$.  We have also calculated several models with
different $\Mdot$-values ($\Mdot_0=10^{-2},10^{-4},10^{-5}\Mdotc$) and
found that the time evolution of the disks is basically similar.

The plausible reason for the wave reflection is as follows: The incident
wave is a thermal mode (Kato et al.  1996) derived by using the short
wavelength approximation, the essential point of which is that the ambient
gas can be regarded homogeneous to the perturbation.  When the thermal mode
reaches the innermost region, where the ambient gas is no longer
homogeneous, it cannot exist as a pure thermal mode and evolves into a
mixture of thermal and acoustic modes.  The thermal mode and the acoustic
mode propagating inward are swallowed by central black hole and only the
acoustic mode propagating outward can escape, which is the reflected wave
itself.

Finally, we calculated the total disk luminosity, $\Ld = \int 2\pi r \Qrad
dr$, and depict the light curve in Figure 3.  We see a broadly peaked,
(time) symmetric profile in the light curve that will be observed as an
X-ray shot.  Using {\it Ginga} data, Negoro, Miyamoto, \& Kitamoto (1994)
directly obtained a mean time profile of the shots by superposing a number
of shots aligning their peaks; it was sharply peaked and is rather (time)
symmetric.  We here demonstrate that such a symmetric light variation can
be reproduced by the advection-dominated disk model.  Although the initial
perturbation amplitude was only $\sim 20$\%, the disk luminosity changed by
a factor of $\sim 60$\%.  Perturbations with amplitudes of 20 -- 30\% are
sufficient to reproduce observed light variations.  We note here that our
prime result of symmetric light curve is a result of wave reflection itself
rather than growth of the wave to the shock wave, since the most of all
luminosity comes from the region inside the radius of $10\rg$ while the
wave grows to the shock wave at the radius larger than $20\rg$.

Another noteworthy feature of X-ray shots is spectral hardening.  The
spectrum is softer than the source's avrage before the peaks of individual
X-ray shots.  It then suddenly becomes harder just after the peak.  (Negoro
et al.  1994) The sudden temperature rise at the innermost region due to
the accostic wave reflection can explain this spectral hardening.  Note
also that temperature of the incident thermal mode is lower than the steady
state because of efficient cooling, which will explain the softer spectrum
before the peak.

In conclusion, the observed X-ray variability can be explained, if the disk
is optically thin and advection-dominated.  Note that the optically thick,
advection-dominated disks are likely to undergo relaxation oscillations
(Honma, Matsumoto, \& Kato 1991).

Although we have assumed the disk to be geometrically thin, this assumption
may not be perfectly satisfied.  Even so, however, Narayan \& Yi (1995)
have shown that the height integrated solutions are good representation of
nearly spherical flows, if one interprets the height integration as a
spherical average.  Hence it seems reasonable to believe that our results
show (at least qualitatively) the properties of time evolution of optically
thin advection dominated accretion flows.  However, two-dimensional
simulations (in the radial and vertical directions) are urgently needed to
confirm our present findings.  When two-dimensional flow is allowed,
circulating flow or turbulence may appear, since the flow speeds in the
equatorial plane and in the disk surface layer may differ.

\noindent{\it 3.2. ~ Origins of X-ray fluctuations}

Several independent mechanisms seem to be involved for producing the
observed X-ray variabilities: (1) generation of disturbances in smooth
accretion flow, (2) formation of X-ray emitting blobs from the disturbance,
and (3) production of complex variability light curves, as are observed, by
superpositions of X-ray shots.  What we have discussed so far concerns the
second step; the thermal instability provides a mechanism to create an
X-ray shot from a disturbance, once the advection-dominated disk is
perturbed somehow.

As to the thrid step above, we note that the observed X-ray fluctuations
are made up of plenty of individual peaks or shots with different peak
intensities.  The peak intensities are smoothly distributed from large ones
to smaller ones, roughly showing a power-law or exponential distribution
(Negoro et al.  1995).  This can be understood if the disk is in a
self-organized critical (SOC) state (see Takeuchi et al.  1996).

The smooth distribution in the released energy of one shot suggests that
the ignition radii, at which a disturbance is given, and initial
perturbation amplitudes are also distributed smoothly from the smaller one
to the larger.  The maximum radius can be determined from the time duration
of the long X-ray shots.  The accretion timescale is roughly given by the
free-fall timescale, which is
$$     \tff \sim \Bigl({r^3\over GM}\Bigr)^{1/2}
           \simeq 4 \Bigl({r\over10^3\rS}\Bigr)^{3/2}
                      \Bigl({M\over 10\Ms}\Bigr) s.     \eqno(6)$$
The duration of $\sim$ 10 s for large X-ray shots corresponds to the free
fall timescale at $\sim 10^3 \rg$ for $M \sim 10\Ms$.  We thus understand
that the maximum radius, at which shots are created, is of the order of
$\sim (1 - 2) \times 10^3\rg$.  This number is a good agreement with the
radius separating the outer, standard-type and inner, advection-dominated
portions of the disk, which was estimated from the fitting to the optical
-- soft X-ray spectra of black-hole X-ray transients during the quiescence
(Narayan, McClintock, \& Yi 1996).

Finally, we discuss the possible seeds for assumed disturbances (the first
step above).  The most promising possibility at present is magnetic flares
(Galeev, Rosener, \& Vaiana 1979; Mineshige, Kusunose, \& Matsumoto 1995).
Because of strong differential rotation and rapid accretion, magnetic
fields are at any times being amplified.  Magnetic reconnection is then a
way to release magnetic energy out of the disk when magnetic pressure
becomes comparable to gas pressure (Shakura \& Sunyaev 1973).  We expect
that magnetic flares can occur when magnetic energy is stored up to a
certain level, regardless of the main radiation mechanisms of the disk.
Once a flare occurs, then a disk material at that part will fall onto the
black hole.  We need further theoretical studies on the detailed flare
mechanism to confirm the above scenario.

\bigskip
\bigskip

We would like to thank Professor Shoji Kato for fruitful discussion. This
work is supported in part by a Scientific Research Grant from the Ministry
of Education, Science and Culture (Nos.  06233101, 07247213; S.M.).

\vfill\eject
\leftskip=0.7cm
\parindent=-0.7cm
\parskip=0.0cm
{\bf References}

Abramowicz, M. A., Bao, G., Lanza, A., \& Zhang, X.-H. 1991,
A\&A,  245, 454

Abramowicz, M. A., Chen, X., Kato, S., Lasota, J.-P., \& Regev, O. 1995,
ApJ, 438, L37

Galeev, A. A., Rosner, R., \& Vaiana, G. S., 1979, ApJ,{ 229}, 318

Honma, F., Matsumoto, R., \& Kato, S. 1991, PASJ 43, 147

Kato, S., Abramowicz, M. A., \& Chen, X. 1996, PASJ, 48, 67

Lawrence, A., Watson, M. G., Pounds, K. A., \& Elvis, M. 1987,
Nature, 325, 694

Lochner, J. C., Swank, J. H., \& Szymkowiak, A. E. 1991, ApJ,375, 295

Matsumoto, R., Fukue, J., Kato, S., Okazaki, A. S. 1984, PASJ, 36, 71

McHardy, I. \& Czerny, B. 1987, Nature { 325}, 696

Mineshige, S., Takeuchi, M., \& Nishimori, H. 1994, ApJ,  435, L125

Mineshige, S., Kusunose, M., \& Matsumoto, R. 1995, ApJ, 445, L43

Miyamoto, S., Kitamoto, S., Iga, S., Negoro, H., \& Terada, K. 1992,
ApJ, 391, L21

Miyamoto, S., Kimura, K., Kitamoto, S., Dotani, T., \& Ebisawa, K. 1991,
ApJ,{ 383}, 784

Narayan, R., McClintock, J., \& Yi, I. 1996, ApJ, Feb 1.

Narayan, R., Yi, I. 1995, ApJ, 452, 710

Negoro, H., Kitamoto, S., Takeuchi, M., \& Mineshige, S. 1995,
  ApJ,  452, L49

Negoro, H., Miyamoto, S., \& Kitamoto, S. 1994, ApJ,  {423}, L27

Oda, M. 1977, Space Sci. Rev., {20}, 757

Piran, T. 1978, ApJ, 221, 652

Shakura, N. I., \& Sunyaev, R. A. 1973, A\&A, 24, 337

Shapiro, S. L., Lightman, A. P., \& Eardley, D. M. 1976, ApJ, 204, 187

Takeuchi, M., Mineshige, S., \& Negoro, H. 1995, PASJ, 47, 617

Terrell, N. J. 1972, ApJ, {174}, L35

\bigskip
{Fig.  1.~} Time evolution of the optically thin, advection-dominated disks
under a density disturbance applied at the outer parts.  The disturbance in
the surface density distribution propagates inward and is reflected at the
innermost parts as an outgoing acoustic wave.  The surface density $\Sigma$
is shown in unit of $g cm^{-2}$.

\bigskip
{Fig.  2.~} The time variations in the radial Mach-number profile ({\it
top}) and radial integrated-pressure profile ($W$, {\it bottom}) in units
of $g/s^{2}$ after the wave reflection.  The elapsed times are 600
(indicated by ``i''), 800, 900, 1000, 1100, and 1200 (``f'') in units of
$c/\rg=10^{-4}$s.  The dotted line represents the initial steady state
(without disturbances).  The acoustic wave emerges and evolves to become a
shock wave.

\bigskip
{Fig.  3.~} The light variation from the advection-dominated disk under the
density disturbance.  The units of time and the disk luminosity are
$c/\rg=10^{-4}$s for $M = 10\Ms$.  The rather symmetric time profile is due
to the wave reflection (cf.  Fig.  1).  The time duration of one X-ray shot
is of the order of the accretion timescale.

\vfill\supereject
\end